\documentclass[aps,prd,onecolumn,groupedaddress,showpacs,nofootinbib,amssymb, preprint]{revtex4}
\usepackage[dvips]{graphicx}
\usepackage{amssymb}
\usepackage{amsmath}
\usepackage{graphicx,color}
\usepackage{amsfonts}
\usepackage{bm}
\usepackage{cancel}
\usepackage{comment}
\usepackage{hyperref}
\usepackage{ulem}

\newcommand{\e}{\mathrm{e}}

\preprint{KEK-TH-2645, KEK-Cosmo-0353}

\allowdisplaybreaks[4]

\begin{document}

\tolerance=5000

\title{Improving Mimetic Gravity with Non-trivial Scalar Potential: \\
Cosmology, Black Holes, Shadow and Photon Sphere}

\author{Shin'ichi~Nojiri$^{1,2}$\,\thanks{nojiri@gravity.phys.nagoya-u.ac.jp},
Sergei~D.~Odintsov$^{3,4}$\,\thanks{odintsov@ieec.uab.es}}
\affiliation{
$^{1)}$ KEK Theory Center, Institute of Particle and Nuclear Studies, 
High Energy Accelerator Research Organization (KEK), Oho 1-1, Tsukuba, Ibaraki 305-0801, Japan \\
$^{2)}$ Kobayashi-Maskawa Institute for the Origin of Particles
and the Universe, Nagoya University, Nagoya 464-8602, Japan \\
$^{3)}$ ICREA, Passeig Lluis Companys, 23, 08010 Barcelona, Spain \\
$^{4)}$ Institute of Space Sciences (ICE, CSIC) C. Can Magrans s/n, 08193 Barcelona, Spain}

\begin{abstract}

It is not easy to treat the spacetime with horizon(s) in the standard mimetic gravity. 
The solution to this problem has been presented in Phys.~Lett.~B 830 (2022), 137140, where it was suggested to modify the Lagrange multiplier constraint. 

In this paper, by using the improved formulation, we investigate the cosmology and black holes in mimetic gravity with scalar potential 
and in the scalar mimetic $F(R)$ gravity. 
The inflationary era and dark energy epoch for the above theories are presented as specific examples from the general reconstruction scheme which permits 
to realise any universe expansion history via the choice of the corresponding scalar potential or function $F(R)$. 
Two black hole solutions including the Schwarzschild and Hayward ones are constructed.
The shadow and the radius of the photon sphere for the above black holes are found. 
The explicit confrontation of the black hole shadow radius with the observational bounds from M87$^*$ and Sgr A$^*$ objects is done. 
It is demonstrated that they do not conflict with Event Horizon Telescope observations. 
\end{abstract}

\maketitle

\section{Introduction}\label{SecI}

The mimetic gravity theory proposed in \cite{Chamseddine:2013kea} has an extra conformal degree of freedom to the gravitational field 
although the degree of freedom is not dynamical.
The additional degree of freedom is expressed by a scalar field $\phi$ and it plays a role of dark matter. 
Hence, it gives the possibility of including dark matter in a geometric way.
It is interesting that mimetic gravity may describe realistic cosmology when the potential of the mimetic scalar is included \cite{Chamseddine:2014vna}. 
Some generalisation of mimetic gravity is mimetic $F(R)$ gravity proposed in \cite{Nojiri:2014zqa}. 
It turns out that within mimetic $F(R)$ gravity it is possible to unify the inflation with dark energy \cite{Nojiri:2017ygt, Nojiri:2016vhu} while dark matter enters via the mimetic scalar.
Different aspects of cosmology and black holes in mimetic gravity have been also studied in \cite{Mirzagholi:2014ifa, Leon:2014yua, Momeni:2015gka, Myrzakulov:2015qaa, Astashenok:2015haa, 
Arroja:2015wpa, Rabochaya:2015haa, Myrzakulov:2015kda, Cognola:2016gjy, Odintsov:2016oyz, Oikonomou:2016pkp, Firouzjahi:2017txv, Hirano:2017zox, Vagnozzi:2017ilo, 
Takahashi:2017pje, Gorji:2017cai, Dutta:2017fjw, Nashed:2018qag, Odintsov:2018ggm, Casalino:2018tcd, Ganz:2018mqi, Solomon:2019qgf, Gorji:2019ttx, Khalifeh:2019zfi, 
Rashidi:2020jao, Gorji:2020ten, Kaczmarek:2021psy, Benisty:2021cin, Nashed:2021ctg, Nashed:2021hgn, Domenech:2023ryc, Nashed:2023jdf, Kaczmarek:2023qmq}, 
for general review see \cite{Sebastiani:2016ras}. 
It is also interesting that mimetic gravity formulation is based on the use of the Lagrange multiplier constraint which was earlier discussed in the context of 
 dark energy epoch in refs.\cite{Lim:2010yk, Gao:2010gj, Capozziello:2010uv}. 
Note that some difficulties may occur when we consider the spacetime with the horizon as in black holes in mimetic gravity. 
As the signature of the metric changes at the horizon, the spacelike (timelike) vector $\partial\phi$ becomes timelike (spacelike), 
the continuity of the scalar field $\phi$ at the horizon becomes highly non-trivial. 
Then even for the static spacetime, the scalar field varies with time. 
This problem has been solved in \cite{Nojiri:2022cah}, where a function $\omega(\phi)$ of the scalar field $\phi$ is introduced. 
This function changes its signature between the region inside the horizon and the region outside the horizon. 
In each of the regions, the function $\omega(\phi)$ is absorbed into the redefinition of the scalar field $\phi$ and we obtain 
the standard mimetic gravity but the redefinition is not continuous at the horizon. 
Hence, the scalar field is smoothly connected at the horizon due to the use of $\omega(\phi)$.

In Ref.~\cite{Gorji:2020ten}, beside the black hole solution where the mimetic scalar field $\phi$ depends on time, the solution with naked singularity 
has been found. 
Recently in the remarkable work \cite{Khodadi:2024ubi}, it was indicated that 
the Event Horizon Telescope observations~\cite{EventHorizonTelescope:2019dse} rule out the compact objects in simplest mimetic gravity 
where only the Lagrange multiplier constraint is included but no scalar potential of mimetic field presents.\footnote{
\color{blue} GW170817 event also constrains the model as in the case of higher-order expansion of mimetic gravity model~\cite{Sharafati:2021egk}. 
}
This was based on the investigation of the black hole shadow~\cite{Held:2019xde, Perlick:2021aok, Chen:2022scf} 
for two classes of spherically symmetric spacetime in baseline mimetic gravity which were found to be pathological. 
For one of the solutions with naked singularity, the shadow is not cast while for the second class solution of the black hole, where the scalar field is time-dependent, 
the radius of the shadow was too small. 
Based on these observations, it was concluded that mimetic gravity under consideration cannot serve as a realistic candidate for dark energy due to the lack of compact objects there.
However, as we show in this paper it turns out that such a conclusion is true in only the simplest mimetic gravity. 

In this paper, following the formulation proposed in \cite{Nojiri:2022cah}, we construct cosmological and black hole (BH) solutions in the framework 
of the scalar mimetic gravity and $F(R)$ gravity with the account of the potential of mimetic scalar as in \cite{Chamseddine:2014vna, Nojiri:2014zqa}. 
For one of the obtained BH solutions, there appears the singular surface beside the horizon but the solution can have a large radius of the black hole shadow 
and therefore the singular surface could not be observed by far observers. 
Furthermore, even if the Arnowitt-Deser-Misner (ADM) mass vanishes in this model, there appears a photon sphere, 
which might be found by the observations of the Event Horizon Telescope. 
We also consider an $F(R)$ gravity extension of the scalar mimetic gravity as in \cite{Nojiri:2014zqa}. 
In the framework of the scalar mimetic $F(R)$ gravity, we explicitly construct a model, where the Hayward black hole~\cite{Hayward:2005gi} is a solution. 
The Hayward black hole is a regular black hole with two horizons. 
The radius of the black hole shadow becomes smaller compared with that in the Schwarzschild black hole with the same ADM mass. 
Therefore the radius might be observed in future observations. 
We also construct the inflationary and dark energy cosmology in the above models.

In the next section, we clarify the problem in the standard mimetic gravity when spacetime with a horizon is considered. 
After that, it is shown how this problem can be solved. 
In Section~\ref{SecIII}, we consider the cosmology in the mimetic gravity with scalar potential and mimetic $F(R)$ gravity. 
It is shown how the arbitrary expansion history of the universe, including inflation and the dark energy era, can be realised in the formulation. 
In Section~\ref{SecIV}, we investigate the static and spherically symmetric spacetime in the above theories. 
Due to the improved formulation, the same theory can describe both the static spherically-symmetric spacetime and cosmology. 
The Schwarzschild black hole and Hayward black hole are constructed for mimetic gravity with non-zero scalar potential and for mimetic $F(R)$ gravity, respectively. 
The radii of the photon sphere and black hole shadow for such black holes are constructed. 
It is demonstrated that corresponding compact objects are not small, they pass observational bounds and may be visible in EHT observations so such mimetic gravities are fully realistic theories. 
Nevertheless, only future observations may distinguish between different modifications of General Relativity. 
The last section is devoted to the summary and discussions.

\section{Scalar mimetic gravity}\label{SecII}

When there exists a horizon, the mimetic gravity theory ~\cite{Chamseddine:2013kea} becomes inconsistent. 
The problem can be solved as in \cite{Nojiri:2022cah}. 
In this section, after a brief review of the model in \cite{Nojiri:2022cah}, we consider mimetic gravity and the mimetic $F(R)$ gravity with non-trivial scalar potential. 

We consider the mimetic gravity~\cite{Chamseddine:2013kea} with scalar potential. 
Here we briefly review the model of \cite{Nojiri:2022cah}
In the mimetic gravitational theory, the conformal degree of freedom of the metric is separated by introducing a relation 
between the auxiliary metric $\bar{g}^{\alpha \beta}$, the physical metric $g_{\alpha \beta}$, and a mimetic field $\phi$ 
as follows, 
\begin{align}
\label{trans1}
g_{\alpha\beta}=\mp \left(\bar{g}^{\mu \nu} \partial_\mu \phi\partial_\nu \phi \right) \bar{g}_{\alpha\beta}\, .
\end{align}
Eq.~(\ref{trans1}) has the scale invariance $\bar{g}_{\mu\nu}\to \e^\sigma \bar{g}_{\mu\nu}$ with a parameter $\sigma$.
Equation~(\ref{trans1}) yields that the mimetic scalar field $\phi$ satisfies 
\begin{align}
\label{trans2}
g^{\alpha \beta}\partial_\alpha \phi \partial_\beta \phi= \mp 1\,.
\end{align}
The mimetic constraint (\ref{trans2}) is not consistent with the black hole geometry with the horizon(s). 
In order to solve this problem, we modify the constraint (\ref{trans2}) by introducing a function $\omega(\phi)$ 
as follows \cite{Nojiri:2022cah}, 
\begin{align}
\label{trans3}
\omega(\phi) g^{\alpha \beta}\partial_\alpha \phi \partial_\beta \phi= \mp 1\, , 
\end{align}
which is locally equivalent to (\ref{trans2}) as we will see soon 
but $\omega(\phi)$ plays an important role when crossing the horizon. 
We should also note that by the constraint (\ref{trans3}), Eq.~(\ref{trans1}) can be rewritten as 
$g_{\alpha\beta}=\mp \omega(\phi) \left(\bar{g}^{\mu \nu} \partial_\mu \phi\partial_\nu \phi \right) \bar{g}_{\alpha\beta}$ 
and it is clear that there remains the scale invariance $\bar{g}_{\mu\nu}\to \e^\sigma \bar{g}_{\mu\nu}$. 

Because we consider the black hole spacetime, we choose the plus sign in (\ref{trans2}) as follows,
\begin{align}
\label{lambdavar0}
g^{\rho \sigma}\partial_\rho\phi \partial_\sigma\phi= 1\, .
\end{align}
We consider the static and spherically symmetric spacetime
with the following line element,
\begin{align}
\label{metD}
ds^2= - \e^{2\nu(r)} dt^2 + \e^{2\eta(r)} dr^2 + r^2 d{\Omega_2}^2\, .
\end{align}
Here $d{\Omega_2}^2$ is the line element of the two-dimensional unit sphere.
If we may also assume $\phi=\phi(r)$, the mimetic constraint (\ref{lambdavar0}) has the following form
\begin{align}
\label{cnstrnt1}
\e^{-2\eta(r)} \left( \phi' \right)^2 = 1 \, .
\end{align}
The equation has no solution if $\e^{-2\eta(r)}$ is negative, $\e^{-2\eta(r)}<0$.
In the case of black hole geometry, $\e^{2\nu(r)}$ vanishes and changes its signature at the horizon.
In order to avoid the curvature singularity, $\e^{2\eta(r)}$ must vanish at the horizon. 
This tells that the mimetic theory with the constraint (\ref{lambdavar0}) cannot realize the black hole geometry with the horizon(s) 
if the solution is static and $\phi$ only depends on $r$. 

In order to avoid the above problem, we may change the mimetic constraint in (\ref{lambdavar0}), a bit different as in (\ref{trans3}). 
If $\omega(\phi)$ is positive, we may define a scalar field $\tilde\phi$ by $\tilde\phi = \int d\phi \sqrt{\omega(\phi)}$,
the constraint (\ref{trans3}) is reduced to the form of (\ref{lambdavar0}),
\begin{align}
\label{lambdavar2}
g^{\rho \sigma}\partial_\rho \tilde\phi \partial_\sigma \tilde\phi= 1\, .
\end{align}
The signature of $\omega(\phi)$ can be, however, changed in general.
If we may also assume $\phi=\phi(r)$ and the spacetime is given by (\ref{metD}),
instead of (\ref{cnstrnt1}), the constraint (\ref{trans3}) has the following form,
\begin{align}
\label{cnstrnt2}
\e^{-2\eta(r)} \omega(\phi) \left( \phi' \right)^2 = 1 \, .
\end{align}
Then for a solution of $\phi$ where $\omega(\phi)$ is positive when $\e^{-2\eta(r)}$ is positive and $\omega(\phi)$ is negative when $\e^{-2\eta(r)}$ is negative,
the constraint (\ref{cnstrnt2}) is consistent even inside the horizon. 
When $\omega(\phi)$ is negative, if we define a scalar field $\hat\phi$ by $\hat\phi=\int d\phi \sqrt{-\omega(\phi)}$, instead of (\ref{lambdavar2}), we obtain
\begin{align}
\label{lambdavar2BB}
g^{\rho \sigma}\partial_\rho \hat\phi \partial_\sigma \hat\phi= - 1\, ,
\end{align}
which corresponds to $-$ signature in (\ref{trans2}). 
Therefore we find that by introducing $\omega(\phi)$, we can treat both signatures $\mp$ in (\ref{trans2}) in a unified way by using a single model. 

As an example, consider a simple case where
\begin{align}
\label{ex1}
\omega(\phi) = \frac{1}{\phi} \, .
\end{align}
Near the horizon, $\e^{-2\eta(r)}$ in (\ref{metD}) should behave as
\begin{align}
\label{ex2}
\e^{-2\eta(r)}(r) \sim b_0 \left( r - r_\mathrm{h} \right) \, .
\end{align}
Here $r_\mathrm{h}$ is the radius of the horizon and $b_0$ is a positive constant.
Then a solution of (\ref{cnstrnt2}) with (\ref{ex1}) is given by
\begin{align}
\label{ex3}
\phi \sim \frac{r - r_\mathrm{h}}{b_0} \, .
\end{align}
Then the scalar $\phi$ and therefore $\omega(\phi)$ change the sign at the horizon and
Eq.~(\ref{cnstrnt2}) is consistent even inside the horizon.

In case there are several horizons, the problem might not be solved only by the choice in (\ref{ex1}). 
As a way to solve the problem in this case, one may choose 
\begin{align}
\label{ex4}
\omega(\phi) = \e^{2\eta(r=\phi)} \, .
\end{align}
In this case, the solution of (\ref{cnstrnt2}) is simply given by 
\begin{align}
\label{ex5}
\phi = r \, .
\end{align}
Therefore it is clear that the problem is solved by the choice of (\ref{ex4}). 
This choice may, however, look rather artificial because it looks like we have assumed the solution from the beginning. 
Anyway, the possibility of the choice (\ref{ex4}) shows that a model gives the solution of Eq.~(\ref{cnstrnt2}). 

\subsection{Scalar mimetic gravity based on Einstein's gravity}\label{SecIIA}

We now consider the action of the mimetic gravity that has the Lagrange multiplier $\lambda (\phi)$ 
and mimetic potential $V(\phi )$ based on Einstein's gravity as follows, 
\begin{align}
\label{actionmimeticfraction}
S=\frac{1}{2\kappa^2}\int \mathrm{d}x^4\sqrt{-g}\left\{ R
 - V(\phi)+\lambda \left(\omega(\phi) g^{\mu \nu}\partial_\mu\phi\partial_\nu\phi
 - 1\right)\right\}+S_\mathrm{matt}\, ,
\end{align}
where $\kappa^2$ is the Einstein gravitational constant which in the relativistic units equals $\kappa^2=8\pi$ and $S_\mathrm{matt}$
is the action of matter. 
The variation of the action (\ref{actionmimeticfraction}) with respect to the metric tensor $g_{\mu \nu}$, yield the equations corresponding to the Einstein equation, 
\begin{align}
\label{aeden}
0=R_{\mu \nu} -\frac{1}{2} g_{\mu \nu} R 
+\frac{1}{2}g_{\mu \nu}\left\{\lambda \left( \omega(\phi) g^{\rho\sigma}\partial_\rho\phi\partial_\sigma\phi - 1\right) -V(\phi)\right\}
 -\lambda \omega(\phi)\partial_\mu\phi \partial_\nu\phi +8\pi T_{\mu \nu} \, . 
\end{align}
Here $T_{\mu \nu}$ is the energy-momentum tensor of matter.
On the other hand, the variation of the action with respect to the mimetic scalar $\phi$ gives 
\begin{align}
\label{scalvar}
\lambda \omega'(\phi) g^{\rho\sigma}\partial_\rho\phi\partial_\sigma\phi
+ 2\nabla^\mu (\lambda\omega(\phi) \partial_\mu\phi) +V'(\phi)=0\, ,
\end{align}
where the ``prime'' or ``$'$'' means the differentiation with respect to the scalar field $\phi$.
Finally, the variation of the action (\ref{actionmimeticfraction}) with respect to the Lagrange multiplier $\lambda$, gives 
the constraint Eq.~(\ref{trans3}). 
Note, the scalar field equation (\ref{scalvar}) can be obtained from (\ref{aeden}) and (\ref{trans3}) when $T_{\mu\nu}=0$ or
by using the conservation law.
Therefore we do not use (\ref{scalvar}) hereafter.
We should also note that when $V(\phi)$ is a constant and $\lambda=0$, Eq.~(\ref{scalvar}) is satisfied and Eq.~(\ref{aeden}) 
reduces to the standard Einstein equation with a cosmological constant $\Lambda=V$, 
\begin{align}
\label{aedenE}
0=R_{\mu \nu} -\frac{1}{2} g_{\mu \nu} R +\frac{1}{2}g_{\mu \nu}\Lambda \, . 
\end{align}
Therefore any vacuum solution of the Einstein gravity like (anti-)de Sitter--Schwarzschild spacetime or (anti-)de Sitter--Kerr spacetime 
is a solution of the model given by (\ref{actionmimeticfraction}). 

\subsection{Scalar mimetic $F(R)$ gravity}\label{SecIIB}

Let us now consider the $F(R)$ gravity extension of the model (\ref{actionmimeticfraction}) as in \cite{Nojiri:2014zqa}, 
\begin{align}
\label{actionmimeticfractionFR}
S=\frac{1}{2\kappa^2}\int \mathrm{d}x^4\sqrt{-g}\left\{ F(R)
 - V(\phi)+\lambda \left(\omega(\phi) g^{\mu \nu}\partial_\mu\phi\partial_\nu\phi
 - 1\right)\right\}+S_\mathrm{matt}\, ,
\end{align}
By the variation of the action (\ref{actionmimeticfractionFR}) with respect to the metric tensor $g_{\mu \nu}$, 
we obtain the equation corresponding to (\ref{aeden})
\begin{align}
\label{aedenFR}
0=&\, - \frac{1}{2}g_{\mu\nu} F + R_{\mu\nu} F_R + g_{\mu\nu} \Box F_R - \nabla_\mu \nabla_\nu F_R \nonumber \\
&\, +\frac{1}{2}g_{\mu \nu}\left\{\lambda \left( \omega(\phi) g^{\rho\sigma}\partial_\rho\phi\partial_\sigma\phi - 1\right) -V(\phi)\right\}
 -\lambda \omega(\phi)\partial_\mu\phi \partial_\nu\phi + \kappa^2 T_{\mu \nu} \, . 
\end{align}
Here $F_R \equiv \frac{dF(R)}{dR}$. 
The variation with respect to the scalar field $\phi$ gives (\ref{scalvar}), and the variation with respect to the Lagrange multiplier $\lambda$ also 
gives (\ref{trans3}). 
Even in the case of the $F(R)$ gravity, Eq.~(\ref{scalvar}) can be obtained from (\ref{aedenFR}) and (\ref{trans3}) and we do not use Eq.~(\ref{scalvar}). 

When $\lambda=0$ and $V$ is a constant, $V=\Lambda$, if we assume that the Ricci curvature is covariantly constant, that is, $R_{\mu\nu}=\frac{1}{4} R g_{\mu\nu}$, 
Eq.~(\ref{aedenFR}) reduces to an algebraic equation for the curvature, 
\begin{align}
\label{aedenFRcc}
0=&\, - 2 F + R F_R - 2 \Lambda\, . 
\end{align}
If the above equation has a real number solution $R=R_0$, when $R_0=0$, the Schwarzschild spacetime and the Kerr spacetime are solutions of this model 
and when $R_0\neq 0$, the (anti-)de Sitter--Schwarzschild spacetime and the (anti-)de Sitter--Kerr spacetime are solutions. 

\section{Cosmological solutions}\label{SecIII}

Before considering spherically symmetric spacetime, we briefly discuss the cosmological solutions. 
A merit considering the model in (\ref{actionmimeticfraction}) with $\omega(\phi)$ is that we can treat both spherically symmetric spacetime and 
in the spatially flat Friedmann-Lema\^{i}tre-Robertson-Walker (FLRW) spacetime, where the metric is given by 
\begin{align}
\label{FLRW}
ds^2 = - dt^2 + a(t)^2 \sum_{i=1,2,3} \left( dx^i \right)^2 \, .
\end{align}
Here $t$ is the cosmological time and $a(t)$ is called a scale factor. 
Then the vanishing components of the connections and the curvatures are given by\footnote{ We use the following convention for the
curvatures and connections:
\begin{align}
R=&\, g^{\mu\nu}R_{\mu\nu} \, , \quad R_{\mu\nu} = R^\lambda_{\
\mu\lambda\nu} \, , \quad R^\lambda_{\ \mu\rho\nu} =
-\Gamma^\lambda_{\mu\rho,\nu} + \Gamma^\lambda_{\mu\nu,\rho} -
\Gamma^\eta_{\mu\rho}\Gamma^\lambda_{\nu\eta} +
\Gamma^\eta_{\mu\nu}\Gamma^\lambda_{\rho\eta} \, ,\nonumber \\
\Gamma^\eta_{\mu\lambda} =&\, \frac{1}{2}g^{\eta\nu}\left(
g_{\mu\nu,\lambda} + g_{\lambda\nu,\mu} - g_{\mu\lambda,\nu}
\right)\, . \nonumber 
\end{align}
} 
\begin{align}
\label{E2}
& \Gamma^t_{ij}=a^2 H \delta_{ij}\, ,\quad \Gamma^i_{jt}=\Gamma^i_{tj}=H\delta^i_{\ j}\, , 
\quad \Gamma^i_{jk}=\tilde \Gamma^i_{jk}\, ,\nonumber \\
& R_{itjt}=-\left(\dot H + H^2\right)a^2\delta_{ij}\, ,\quad 
R_{ijkl}= a^4 H^2 \left(\delta_{ik} \delta_{lj} - \delta_{il} \delta_{kj}\right)\, ,\nonumber \\
& R_{tt}=-3\left(\dot H + H^2\right)\, ,\quad R_{ij}= a^2 \left(\dot H + 3H^2\right)\delta_{ij}\, ,\quad R= 6\dot H + 12 H^2\, , \nonumber \\
& \mbox{other components}=0\ .
\end{align}
Here the Hubble rate $H$ is defined by $H=\frac{\dot a}{a}$ and ``dot'' or ``$\dot\ $'' is the derivative with respect to the cosmological time $t$. 
We now assume the mimetic scalar field $\phi$ only depends on $t$. 

\subsection{Cosmology in scalar mimetic gravity}\label{SecIIIA}

We now consider the cosmology in the framework of scalar mimetic gravity (\ref{actionmimeticfraction}). 

The $(t,t)$ and $(i,j)$ components of (\ref{aeden}) have the following forms, 
\begin{align}
\label{FLRW1}
0= 3H^2 + \frac{1}{2} V (\phi) - \lambda \omega (\phi) \left( \dot\phi \right)^2 - \kappa^2 \rho \, , \quad 
0= -2 \dot H - 3 H^2 - \frac{1}{2} V (\phi) - \kappa^2 p \, .
\end{align}
Here $\rho$ and $p$ are the energy density and the pressure of matter. 

On the other hand, Eq.~(\ref{trans3}) has the following form, 
\begin{align}
\label{lambdavarFLRW}
- \omega(\phi) \left( \dot \phi \right)^2 = 1\, ,
\end{align}
For the choice (\ref{ex1}), a solution of (\ref{lambdavarFLRW}) is given by 
\begin{align}
\label{FLRW2}
\phi= - \frac{t^2}{4} \, .
\end{align}
Then as in the original paper on the mimetic gravity~\cite{Chamseddine:2013kea}, the mimetic scalar field plays the role of dark matter. 

As an example, when there is no matter, $\rho=p=0$ and the potential $V(\phi)$ vanishes, $V(\phi)=0$, the equations (\ref{FLRW1}) have the following form, 
\begin{align}
\label{FLRW3}
0= 3H^2 + \lambda \, , \quad 0= -2 \dot H - 3 H^2 \, .
\end{align}
Here we have used (\ref{FLRW2}). 
A solution is given by 
\begin{align}
\label{FLRW4}
H=\frac{\frac{2}{3}}{t} \, , \quad \lambda = - \frac{\frac{4}{3}}{t^2} \, .
\end{align}
This is nothing but the cosmological expansion driven by the dark matter. 

We may realise any expansion expressed by $H=H(t)$ by adjusting $V(\phi)$. 
Here we neglect the contribution from the matter by putting $\rho=p=0$, again. 
Then the second equation (\ref{FLRW1}) gives the time dependence of $V(\phi)$, $V=V(t)$, as follows, 
\begin{align}
\label{FLRW5}
V(t)= - 4 \dot H(t) - 6 H(t)^2\, .
\end{align}
Because the time dependence of $\phi$ is given by (\ref{FLRW2}), one can express $V$ as a function of $\phi$ 
as $V(\phi)=V\left( t = 2 \sqrt{-\phi} \right)$. 
The first equation of (\ref{FLRW1}) gives the time dependence of $\lambda$\, 
\begin{align}
\label{FLRW6}
\lambda(t) = - 3 H(t)^2 - \frac{1}{2} V(t) = 3H(t)^2 + 2\dot H(t)\, .
\end{align}
Here we used (\ref{FLRW5}). 
Therefore an arbitrary expansion history of the universe can be realized. 

As an example, one may construct a model mimicking the $\Lambda$CDM model without real matter or dark matter. 
In the case of the $\Lambda$CDM model, the Hubble rate is given by
\begin{align}
\label{LCDM2}
H(t) = \frac{2}{3}\alpha \coth \left( \alpha t \right)\, . \quad 
\dot H (t) = - \frac{2\alpha^2}{3 \sinh^2 \left( \alpha t \right)} \, .
\end{align}
Then Eq.~(\ref{FLRW5}) and (\ref{FLRW2}) give, 
\begin{align}
\label{FLRW7}
V= - \frac{8}{3} \alpha^2 \coth^2 \left( \alpha t \right) + \frac{8\alpha^2}{3\sinh^2 \left( \alpha t \right)} 
= \frac{8}{3} \alpha^2 \, .
\end{align}
One also finds the solution of $\lambda$ by using (\ref{FLRW6}), 
\begin{align}
\label{FLRW8}
\lambda(t) = \frac{4}{3} \alpha^2 \coth^2 \left( \alpha t \right) - \frac{4\alpha^2}{3\sinh^2 \left( \alpha t \right)} = \frac{4}{3} \alpha^2\, .
\end{align}
Therefore $V$ plays the role of the cosmological constant $V=\Lambda=\frac{8}{3} \alpha^2$ and the mimetic scalar plays the role of the dark matter. 
The corresponding black hole solution is the Schwarzschild-de Sitter spacetime which will be given later. 

\subsection{Cosmology in scalar mimetic $F(R)$ gravity}\label{SecIIIB}

Let us consider the spatially flat FLRW cosmology (\ref{FLRW}) by using the scalar mimetic $F(R)$ gravity as in \cite{Nojiri:2014zqa}. 
In the spacetime (\ref{FLRW}), by assuming that the mimetic scalar field $\phi$ only depends on $t$, the $(t,t)$ and $(i,j)$ components of (\ref{aedenFR}) have the following forms, 
\begin{align}
\label{Mi34}
0 =& - F(R) + 6 \left( \dot H + H^2 \right) F_R(R) - 6H \frac{d F_R(R)}{dt}
 - \lambda\left( \omega(\phi) {\dot \phi}^2 - 1 \right) + V(\phi) + \kappa^2 \rho\, , \\
\label{Mi35}
0=& F(R) - 2 \left(\dot H + 3H^2 \right) F_R(R) + 2 \frac{d^2 F_R(R)}{dt^2}
+ 4H \frac{d F_R(R)}{dt} - \lambda\left( \omega(\phi) {\dot \phi}^2 + 1 \right) \nonumber \\
&\, - V(\phi) + \kappa^2 p \, .
\end{align}
We also obtain (\ref{lambdavarFLRW}) by the variation of the action with respect to $\lambda$. 
Under the (\ref{ex1}), we obtain (\ref{FLRW2}) as a solution of (\ref{lambdavarFLRW}), which gives $t=2 \sqrt{-\phi}$. 
Then Eqs.~(\ref{Mi34}) and (\ref{Mi35}) reduce to the following forms, 
\begin{align}
\label{Mi34B}
0 =& - F(R) + 6 \left( \dot H + H^2 \right) F_R(R) - 6H \frac{d F_R(R)}{dt}
 - 2 \lambda + V(\phi) + \kappa^2 \rho\, , \\
\label{Mi35B}
0=& F(R) - 2 \left(\dot H + 3H^2 \right) F_R(R) + 2 \frac{d^2 F_R(R)}{dt^2}
+ 4H \frac{d F_R(R)}{dt} - V(\phi) + \kappa^2 p \, .
\end{align}
By combining (\ref{Mi34B}) and (\ref{Mi35B}), we may delete $V(\phi)$ as follows, 
\begin{align}
\label{Mi35C}
0 =& 4 \left( \dot H + H^2 \right) F_R(R) - 2 H \frac{d F_R(R)}{dt} + 2 \frac{d^2 F_R(R)}{dt^2}
 - 2 \lambda + \kappa^2 \left( \rho + p \right)\, .
\end{align}
In the following, by using (\ref{Mi35B}) and (\ref{Mi35C}), we consider the cosmology. 

When $\rho$ and $p$ satisfy any equation of state (EoS), $p=p(\rho)$, if the conservation law 
\begin{align}
\label{cons}
0=\dot\rho + 3 H \left( \rho + p \right)\, , 
\end{align}
is satisfied, $\rho$ and $p$ are expressed as functions of the scale factor $a$, $\rho=\rho(a)$ and $p=p(a)$. 

The expansion history can be expressed by the scale factor $a$ as a function of $t$, $a=a(t)$. 
We consider how one can construct a model realising the expansion history expressed by $a=a(t)$. 
Because $\rho$ and $p$ are expressed as functions of the scale factor $a$, $\rho$ and $p$ can be also expressed by 
functions of $t$, $\rho(t)=\rho\left(a\left(t\right)\right)$ and $p(t)=p\left(a\left(t\right)\right)$. 
Because the scalar curvature is given by $R=12H^2 + 6\dot H$, we find the $t$ dependence of $R$ from $a=a(t)$ via $H(t)=\dot a(t)/a(t)$. 
Then if we properly give the form of $F(R)$, by using (\ref{Mi35C}), we find the solution of $\lambda$ as a function of $t$, 
\begin{align}
\label{Fcoslambda}
\lambda(t) = \left( \dot H(t) + H(t)^2 \right) F_R(R(t)) - H \frac{d F_R(R)}{dt} + \frac{d^2 F_R(R(t))}{dt^2}
+ \frac{\kappa^2}{2} \left( \rho(t) + p(t) \right)\, .
\end{align}
On the other hand, by using (\ref{Mi35B}), we find the $t$-dependence of $V(\phi)$, 
\begin{align}
\label{FcosV}
V(t) =&\, F(R(t)) - 2 \left(\dot H(t) + 3H(t)^2 \right) F_R(R(t)) + 2 \frac{d^2 F_R(R(t))}{dt^2}
+ 4H \frac{d F_R(R(t))}{dt} \nonumber \\
&\, + \kappa^2 p(t) \, .
\end{align}
Because $t=2 \sqrt{-\phi}$, we find the corresponding $V$ as a function of $\phi$, $V(\phi) =V\left( t=2 \sqrt{-\phi}\right)$. 
Therefore for arbitrary expansion history given by $a=a(t)$ and an arbitrary functional form of $F(R)$, we find the corresponding potential $V(\phi)$. 
Inversely, if we start with the functional form of $F(R)$ and the potential $V(\phi)$, the scale factor $a=a(t)$ given first is a solution of the model. 

\subsubsection{Inflation}\label{SecIIIB1}

We may consider the following model proposed in \cite{Nojiri:2024zab} as a model of inflation, 
\begin{align}
\label{Hex1b}
H(t) = \frac{H_0}{1 + \alpha \ln \left( 1 + \e^{\frac{2 H_0}{\alpha} \left( t - t_0 \right)} \right)} \, .
\end{align}
Here $\alpha$ is a positive constant and $t_0$ is a constant corresponding to the time when the inflation ends. 
When $t\ll t_0$, $H$ goes to a constant $H\to H_0$, which corresponds to the inflation. 
When $t\gg t_0$, we find $H\to \frac{1}{2\left(t - t_0 \right)}$, whose behaviour expresses the radiation-dominated universe. 
The following constraints can be satisfied in the model (\ref{Hex1b}) as shown in \cite{Nojiri:2024zab}, 
\begin{align}
\label{index}
n_s = 0.9649 \pm 0.0042\, , \quad r < 0.064\, ,
\end{align}
These constraints were obtained by the Planck 2018 observation. 

Due to 
\begin{align}
\label{derivatives}
\dot H = - \frac{2 H^2}{1 + \e^{-\frac{2 H_0}{\alpha} \left( t - t_0 \right)}} \, ,
\end{align}
one finds 
\begin{align}
\label{Rinf}
R=\frac{12 {H_0}^2 \e^{-\frac{2 H_0}{\alpha} \left( t - t_0 \right)}}
{\left( 1 + \e^{-\frac{2 H_0}{\alpha} \left( t - t_0 \right)} \right)
\left\{1 + \alpha \ln \left( 1 + \e^{\frac{2 H_0}{\alpha} \left( t - t_0 \right)} \right) \right\}^2} \, ,
\end{align}
which gives the $t$ dependence of $F(R)$ for a given functional form of $F(R)$. 

At the end of the inflation, the matter could be generated by the quantum corrections. 
The classical action does not include these effects. 
As in \cite{Nojiri:2024zab}, one may effectively include the effects by modifying the energy density $\rho$ and the pressure $p$, as follows, 
\begin{align}
\label{effrhop}
\rho \to \rho_\mathrm{eff} \equiv \rho + \mathcal{J}_\rho (t) \, , \quad 
p \to p_\mathrm{eff} \equiv p + \mathcal{J}_p(t)\, \, .
\end{align}
Then due to the Bianchi identity, $\rho_\mathrm{eff}$ and $p_\mathrm{eff}$ must satisfy the conservation law as in (\ref{cons}), 
$0=\dot\rho_\mathrm{eff} + 3 H \left( \rho_\mathrm{eff} + p_\mathrm{eff} \right)$. 
Therefore we obtain, 
\begin{align}
\label{conseff}
\dot\rho + 3 H \left( \rho + p \right) = J \equiv - \dot{\mathcal{J}}_\rho - 3 H \left( \mathcal{J}_\rho + \mathcal{J}_p \right) \, .
\end{align}
Eq.~(\ref{conseff}) tells that $J$ plays the role of a source of matter. 
We choose $J$ not to vanish only just after the inflation so that $J$ generates matter at the time. 

Similar to \cite{Nojiri:2024zab}, we choose $\rho$ and $p=\frac{\rho}{3}$ in the case of the radiation, whose EoS parameter is $\frac{1}{3}$, 
\begin{align}
\label{rhop}
\rho=3p= \frac{3H^2 \left( {H_0}^2 - H^2 \right)}{\kappa^2 \left( {H_0}^2 + H^2 \right)} \, .
\end{align}
Then we find $\rho \to 0$ when $t\ll t_0$. 
On the other hand, when $t\gg t_0$, we find the behaviour of $\rho$ as $\rho\to - \frac{3Q}{6\kappa^2} = \frac{3H^2}{\kappa^2}$, as in Einstein's gravity. 
Eqs.~(\ref{Hex1b}) and (\ref{conseff}) give
\begin{align}
\label{J}
J = \frac{12H^3}{\kappa^2\left( {H_0}^2 + H^2 \right)} \left\{ - \frac{{H_0}^4 - 2 {H_0}^2 H^2 - H^4}
{\left( {H_0}^2 + H^2 \right) \left( 1 + \e^{-\frac{2 H_0}{\alpha} \left( t - t_0 \right)} \right)}
+ {H_0}^2 - H^2 \right\} \, .
\end{align}
As we require, $J$ vanishes at the early time $t\ll t_0$ and at the late time $t\gg t_0$. 
By choosing $\mathcal{J}_\rho(t)=0$ in Eq.~(\ref{conseff}), we obtain
\begin{align}
\label{fp}
\mathcal{J}_p (t) = - \frac{4H^2}{\kappa^2\left( {H_0}^2 + H^2 \right)} \left\{ - \frac{{H_0}^4 - 2 {H_0}^2 H^2 - H^4}
{\left( {H_0}^2 + H^2 \right) \left( 1 + \e^{-\frac{2 H_0}{\alpha} \left( t - t_0 \right)} \right)}
+ {H_0}^2 - H^2 \right\} \, .
\end{align}
which may express the quantum generation of the radiation effectively. 

Then we obtain the $t$ dependence of $\lambda$ and $V$ as follows, 
\begin{align}
\label{Fcoslambdainf}
\lambda(t) =&\, \left( \dot H(t) + H(t)^2 \right) F_R(R(t)) - H \frac{d F_R(R)}{dt} + \frac{d^2 F_R(R(t))}{dt^2} 
\nonumber \\
&\, + \frac{\kappa^2}{2} \left( \rho(t) + p(t) + \mathcal{J}_p (t) \right)\, , \\
\label{FcosVinf}
V(t) =&\, F(R(t)) - 2 \left(\dot H(t) + 3H(t)^2 \right) F_R(R(t)) + 2 \frac{d^2 F_R(R(t))}{dt^2}
+ 4H \frac{d F_R(R(t))}{dt} \nonumber \\
&\, + \kappa^2 \left( p(t) + \mathcal{J}_p (t) \right) \, .
\end{align}
By using $t=2\sqrt{-\phi}$, we find $V$ as a function of $\phi$, $V=V(\phi)$. 

\subsubsection{Mimicking $\Lambda$CDM model}\label{SecIIIB2}

We now consider a theory mimicking the $\Lambda$CDM model in (\ref{LCDM2}), where scale factor is given by 
\begin{align}
\label{LCDM1}
a(t) = a_0 \sinh^\frac{2}{3} \left( \alpha t \right)\, ,
\end{align}
where $a_0$ is a positive constant. 
By using (\ref{LCDM2}), we find that the scalar curvature is given by 
\begin{align}
\label{RlCDM}
R (t) = \left( \frac{4}{3} \coth^2 \left( \alpha t \right) + 4 \right) \alpha^2 \, .
\end{align}
Therefore we find the $t$ dependence of $F(R)$ whose functional expression is given. 
We also assume the matter is given by dust, which could be baryonic matter, where
\begin{align}
\label{Ldust}
\rho= \rho_0 a(t)^{-3} = \frac{\rho_0}{{a_0}^3 \sinh^2 \left(\alpha t \right)}\, , \quad p=0\, .
\end{align}
Then for a given functional form of $F(R)$, by using Eqs.~(\ref{Fcoslambda}) and (\ref{FcosV}), we find
\begin{align}
\label{FcoslambdaLCDM}
\lambda(t) =&\, \left( \frac{2}{3} - \frac{2}{9} \coth^2 \left( \alpha t \right) \right) F_R(R(t)) 
 - \frac{2}{3}\alpha \coth \left( \alpha t \right) \frac{d F_R(R)}{dt} + \frac{d^2 F_R(R(t))}{dt^2} \nonumber \\
&\, + \frac{\kappa^2}{2} \frac{\rho_0}{{a_0}^3 \sinh^2 \left(\alpha t \right)} \, , \\
\label{FcosVLCDM}
V(t) =&\, F(R(t)) - \frac{4}{3} \alpha^2 \left( 1 + \coth^2 \left( \alpha t \right) \right) F_R(R(t)) + 2 \frac{d^2 F_R(R(t))}{dt^2} \nonumber \\
&\, + \frac{8}{3}\alpha \coth \left( \alpha t \right) \frac{d F_R(R(t))}{dt} \, .
\end{align}
Here $R(t)$ is given by (\ref{RlCDM}). 
By substituting $t=2\sqrt{-\phi}$ into $V(t)$ in (\ref{FcosVLCDM}), we find the potential $V(\phi)$. 

\subsubsection{Unification of inflation and dark energy}\label{SecIIIB3}

The unification of the inflation and dark energy epochs 
has been achieved in $f(R)$ gravity~\cite{Nojiri:2003ft, Nojiri:2010wj} some time ago. 
In this subsection, we consider the unification in the framework of the scalar mimetic $F(R)$ gravity. 

As in \cite{Nojiri:2024zab}, we assume that the energy density $\rho$ and the pressure $p$ are given by 
\begin{align}
\label{Uni1}
\rho (a) = \frac{a^n}{1 + a^n } \left( \rho_0^\mathrm{radiation} a^{-4} + \rho_0^\mathrm{baryon} a^{-3} \right) \, , \quad 
p (a) = \frac{\rho_0^\mathrm{radiation} a^{n-4} }{3\left( 1 + a^n \right) } \, , 
\end{align}
where $\rho_0^\mathrm{radiation}$ and $\rho_0^\mathrm{baryon}$ are positive constants and $n$ is assumed to be positive integer larger than 4. 
We assume that the matter was generated at the end of the inflation as in (\ref{conseff}). 
The factor $\frac{a^n}{1 + a^n}$ expresses the creation of matter. 
In addition to the radiation $\rho_0^\mathrm{radiation} a^{-4}$, 
we include the baryonic matter $rho_0^\mathrm{baryon} a^{-3}$. 
When $a\ll 1$, one finds $\frac{a^n}{1 + a^n} \sim a^n \to 0$ and therefore $\rho \to 0$, 
and when $a\gg 1$, $\frac{a^n}{1 + a^n} \to 1$ and $\rho \to \rho_0^\mathrm{radiation} a^{-4} + \rho_0^\mathrm{baryon} a^{-3}$, that is, the sum 
of the standard radiation and the baryon. 

By using (\ref{Uni1}), we find the source $J$ of matter in (\ref{conseff}) is given by 
\begin{align}
\label{Juni}
J = \frac{n H \left( \rho_0^\mathrm{radiation} a^{2n -4} + \rho_0^\mathrm{baryon} a^{2n-3} \right) } {\left( 1 + a^n \right)^2} \, .
\end{align}
By choosing $\mathcal{J}_\rho(t)=0$ in Eq.~(\ref{conseff}), again, we obtain
\begin{align}
\label{Jpuni}
\mathcal{J}_p= - \frac{n \left( \rho_0^\mathrm{radiation} a^{2n -4} + \rho_0^\mathrm{baryon} a^{2n-3} \right) } {3\left( 1 + a^n \right)^2} \, .
\end{align}
We should note $J$ and $\mathcal{J}_p$ go to vanish when $a$ is small, $a\ll 1$ or $a$ is large $a\gg 1$. 
This could tell that the end of the inflation occurs when $a\sim 1$ and the generation of matter occurs at that time. 

Similar to the model in \cite{Nojiri:2024zab}, we consider the model that $H$ is given by, 
\begin{align}
\label{Uni2}
H^2 = \frac{{H_0}^2 \left( 1 + \epsilon a^n \right)}{1 + a^n} - \frac{\kappa^2}{3} \rho \, ,
\end{align}
with positive constants $H_0$ and $\epsilon$. 
When $a\ll 1$, the first term in the r.h.s. behaves as $\frac{{H_0}^2 \left( 1 + \epsilon a^n\right) }{1 + a^n} \to {H_0}^2$ when $a\ll 1$, 
which corresponds to the large effective cosmological constant generating inflation. 
On the other hand, when $a\gg 1$, the first term behaves as $\frac{{H_0}^2 \left( 1 + \epsilon a^n\right)}{1 + a^n} \to \epsilon{H_0}^2$, 
which gives the small effective cosmological constant generating the late-time accelerating expansion by choosing $\epsilon$ to be very small. 

Because $H=\frac{\dot a}{a}$, Eq.~(\ref{Uni2}) can be integrated as
\begin{align}
\label{U1}
t = \int da \sqrt{\frac{1 + a^n}{ {H_0}^2 \left( 1 + \epsilon a^n \right) - \frac{\kappa^2}{3} 
\left( \rho_0^\mathrm{radiation} a^{n-4} + \rho_0^\mathrm{baryon} a^{n-3} \right)}} \, .
\end{align}
Eq.~(\ref{U1}) gives $t$ as a function of $a$, which could be algebraically solved with respect to $a$, $a=a(t)$. 
Then we obtain the scalar curvature $R$ as a function of $t$, which also gives $t$-dependence of $F(R)$. 
Then by using (\ref{Fcoslambdainf}) and (\ref{FcosVinf}), we find $t$ dependences of $\lambda$ and $V$, $\lambda=\lambda(t)$ and $V=V(t)$. 
Further by substituting the expression $t=2\sqrt{-\phi}$, we find $V$ as a function of $\phi$, $V=V(\phi)$. 
These are rather general considerations. 
One can confront the theory under discussion with observational bounds to obtain more precise constraints to theory parameters and functions.

\section{Static and spherically symmetric solution}\label{SecIV}

In this section, we consider static and spherically symmetric spacetime in (\ref{metD}). 

By writing ${d\Omega_2}^2=\sum_{i,j=1,2} \tilde g_{ij} dx^i dx^j$ in (\ref{metD}), 
the non-vanishing connections and curvatures are,
\begin{align}
\label{GBv0}
\Gamma^r_{tt}=&\, \e^{-2\eta + 2\nu}\nu' \, ,\quad \Gamma^t_{tr}=\Gamma^t_{rt}=\nu'\, ,\quad 
\Gamma^r_{rr}=\eta'\, ,\quad \Gamma^i_{jk}=\tilde \Gamma^i_{jk}\ ,\quad 
\Gamma^r_{ij}=-\e^{-2\eta}r\tilde g_{ij}\, , \nonumber \\
\Gamma^i_{rj}=&\, \Gamma^i_{jr}=\frac{1}{r}\delta^i_{\ j}\, , \\
\label{GBv}
R_{rtrt}=&\,\e^{2\nu}\left\{\nu'' + \left(\nu' - \eta'\right) \nu'\right\} \, , \quad  
R_{titj}= r\nu'\e^{2(\nu - \eta)} \tilde g_{ij} \, ,\quad  
R_{rirj}= r\eta' \tilde g_{ij} \, ,\nonumber \\
R_{ijkl}=&\, \left(1 - \e^{-2\eta}\right) r^2 \left(\tilde g_{ik} \tilde g_{jl} - \tilde g_{il} \tilde g_{jk} 
\right)\, ,\nonumber \\
R_{tt}=&\, \e^{2\left(\nu - \eta\right)} \left\{ \nu'' + \left(\nu' - \eta'\right)\nu' + \frac{2 \nu'}{r}\right\} \, ,\quad 
R_{rr}= - \left\{ \nu'' + \left(\nu' - \eta'\right)\nu' \right\} + \frac{2 \eta'}{r} \ ,\nonumber \\
R_{ij}=&\, \left[ 1 - \left\{1 + r \left(\nu' - \eta' \right)\right\}\e^{-2\eta}\right] \tilde g_{ij}\, , \nonumber \\
R=&\, 2 \e^{-2\eta}\left[ - \nu'' - \left(\nu'  - \eta'\right)\nu' - \frac{2\left( \nu' - \eta' \right)}{r} + \frac{\e^{2\eta} - 1}{r^2} \right] \, .
\end{align}
In (\ref{GBv0}) and (\ref{GBv}), we denote the derivative with respect to $r$ by ``prime'' or ``$'$''. 

\subsection{Static and spherically symmetric solution in scalar mimetic gravity}\label{SecIVA}

Let us consider the spherically symmetric solution in scalar mimetic gravity in (\ref{actionmimeticfraction}).

Then $(t,t)$, $(r,r)$, and $(i,j)$ components of Eq.~(\ref{aeden}) are given by 
\begin{align}
\label{eq1}
0=&\, \e^{2\left(\nu - \eta \right)} \left( \frac{2\eta'}{r} + \frac{\e^{2\eta} - 1}{r^2} \right) 
+ \frac{1}{2}\e^{2\nu} V \, , \\
\label{eq2}
0=&\, \frac{2\nu'}{r} - \frac{\e^{2\eta} - 1}{r^2} - \frac{\e^{2\eta}}{2} V - \lambda \omega \left(\phi'\right)^2 \, , \\
\label{eq3}
0=&\, r^2 \e^{-2\eta} \left\{ \nu'' + \left(\nu' - \eta' \right) \nu' + \frac{\nu' - \eta'}{r} \right\} - \frac{r^2}{2} V \, .
\end{align}
Here we used (\ref{trans3}), which has the following form, 
\begin{align}
\label{eq4}
\e^{-2\eta} \omega \left( \phi' \right)^2 = 1 \, .
\end{align}
By combining (\ref{eq1}), (\ref{eq2}), and (\ref{eq4}), we obtain 
\begin{align}
\label{eqf1}
0 = 2 \left( \nu' + \eta' \right) - r\lambda \e^{2\eta} \, .
\end{align}
On the other hand, by combining (\ref{eq1}) and (\ref{eq3}), one obtains 
\begin{align}
\label{eqf2}
0 = \nu'' + \left(\nu' - \eta' \right) \nu' + \frac{\nu' + \eta'}{r} + \frac{\e^{2\eta} - 1}{r^2} \, .
\end{align}
Eq.~(\ref{eqf2}) can be rewritten as 
\begin{align}
\label{eqf2B}
0= \e^{-\eta} r \left\{ \left( \e^\nu \right)'' + \left( \frac{1}{r} - \eta' \right) \left(\e^\nu \right)' 
+ \left( \frac{\eta'}{r} + \frac{\e^{2\eta} - 1}{r^2} \right) \e^\nu \right\} \, ,
\end{align}
which is a homogeneous linear differential equation for $\e^\nu$. 

For the Schwarzschild spacetime $\e^{2\eta}= \frac{1}{1 - \frac{r_0}{r}}$, which is a solution of the model, 
Eq.~(\ref{eq4}) has the following form, 
\begin{align}
\label{eq4B1}
\frac{1}{1 - \frac{r_0}{r}} \omega \left( \phi' \right)^2 = 1 \, .
\end{align}
We now choose $\omega$ by (\ref{ex1}). 
Then when $\frac{r_0}{r}<1$, Eq.~(\ref{eq4B1}) has the following form 
\begin{align}
\label{omo1}
\left( \phi^\frac{1}{2} \right)' = \frac{1}{4} \sqrt{ 1 - \frac{r_0}{r}} \, .
\end{align}
Therefore we obtain 
\begin{align}
\label{omo1}
\phi^\frac{1}{2} = \frac{r_0}{8} \left\{ \frac{1}{1 - \sqrt{ 1 - \frac{r_0}{r}}} - \frac{1}{1 + \sqrt{ 1 - \frac{r_0}{r}}} 
+ \ln \frac{1 - \sqrt{ 1 - \frac{r_0}{r}}}{1 + \sqrt{ 1 - \frac{r_0}{r}}} \right\} + C_0
\end{align}
Let us choose the constant of the integration $C_0=0$ so that $\phi$ vanishes at the horizon $r=r_0$. 
On the other hand when $\frac{r_0}{r}>1$, Eq.~(\ref{eq4B1}) has the following form 
\begin{align}
\label{omo1B}
\left( \left( - \phi\right)^\frac{1}{2} \right)' = \frac{1}{4} \sqrt{\frac{r_0}{r} - 1} \, ,
\end{align}
and we obtain 
\begin{align}
\label{omo1B}
\left( - \phi\right)^\frac{1}{2} = \frac{r_0}{8} \left\{ \frac{i}{1 - i \sqrt{ \frac{r_0}{r} - 1}} - \frac{i}{1 + i\sqrt{\frac{r_0}{r}-1}} 
 - i \ln \frac{1 - i\sqrt{ \frac{r_0}{r} - 1}}{1 +i \sqrt{\frac{r_0}{r} - 1}} \right\} + {\tilde C}_0
\end{align}
We choose the constant of the integration ${\tilde C}_0=0$ so that $\phi$ vanishes at the horizon $r=r_0$, again. 

When $V$ is a constant $V=\Lambda$, a solution when $\lambda=0$ is the Schwarzschild-de Sitter spacetime, 
\begin{align}
\label{SdS1}
\e^{2\nu}=\e^{-2\eta}= 1 - \frac{r_0}{r} - \frac{r^2}{l^2}\, , 
\end{align}
with length parameter $l$ defined by $\Lambda=\frac{6}{l^2}$. 
We may rewrite (\ref{SdS1}) by using $r_\pm$ $\left(r_+ > r_- >0\right)$ as follows, 
\begin{align}
\label{SdS2}
\e^{2\nu}=&\, \e^{-2\eta}= 1 - \frac{r_0}{r} - \frac{r^2}{l^2}= - \frac{\left(r_+ + r_- + r \right)\left( r_- - r \right) \left( r_+ - r \right)}{l^2r} \, , \nonumber \\
&\, l^2 = {r_+}^2 + {r_-}^2 + r_+ r_-\, , \quad r_0= \frac{\left(r_+ + r_-\right) r_+ r_-}{{r_+}^2 + {r_-}^2 + r_+ r_-} \, .
\end{align}
The surface $r=r_+$ corresponds to the outer horizon and the surface $r=r_-$ to the inner one. 

When $r_-<r<r_+$, Eq.~(\ref{eq4}) with (\ref{ex1}) gives 
\begin{align}
\label{omo1SdS}
\phi^\frac{1}{2} = \int dr \frac{1}{4} \sqrt{ - \frac{\left(r_+ + r_- + r \right)\left( r_- - r \right) \left( r_+ - r \right)}{l^2r}} \, .
\end{align}
and when $r>r_+$ or $r<r_-$, we find 
\begin{align}
\label{omo1SdS2}
\phi^\frac{1}{2} = - \int dr \frac{1}{4} \sqrt{ \frac{\left(r_+ + r_- + r \right)\left( r_- - r \right) \left( r_+ - r \right)}{l^2r}} \, .
\end{align}
The above integrations are rather tedious. 

\subsubsection{Non-trivial solution when $V=0$}\label{SecIVA1}

We now consider the solution when $V=0$. 
Then Eq.~(\ref{eq1}) can be rewritten as 
\begin{align}
\label{eqs1}
\frac{1}{r} = \frac{2\eta'}{1 - \e^{2\eta}} = - \left( \ln \left( \e^{-2\eta} - 1 \right) \right)' \, ,
\end{align} 
whose solution is 
\begin{align} 
\label{eqs2}
\e^{-2\eta} - 1 = - \frac{r_0}{r} \, ,
\end{align}
with a constant of the integration $r_0$, 
that is, the solution of $\e^{2\eta}$ is that of the Schwarzschild one, 
\begin{align}
\label{eqs3}
\e^{2\eta}= \frac{1}{1 - \frac{r_0}{r}}\, .
\end{align}
Then the solution of Eq.~(\ref{eq4}) is given by (\ref{omo1}) and (\ref{omo1B}). 

Eq.~(\ref{eqf2B}) gives, 
\begin{align}
\label{eqf4}
0 = \left( \e^\nu \right)'' + \frac{1}{2} \left( \frac{\frac{2}{r} - \frac{r_0}{r^2}}{1 - \frac{r_0}{r}} \right) \left(\e^\nu \right)' 
+ \frac{1}{2} \frac{ \frac{r_0}{r^3}}{1 - \frac{r_0}{r}} \e^\nu
\end{align}
The Schwarzschild spacetime $\e^{2\nu}=1 - \frac{r_0}{r}$ is a solution (\ref{eqf4}). 
Then by assuming 
\begin{align}
\label{eqf5}
\e^\nu = \xi(r) \left(1 - \frac{r_0}{r}\right)^\frac{1}{2} \, ,
\end{align}
we find 
\begin{align}
\label{eqf7}
\ln \frac{\xi'(r)}{\xi_0 r_0} = - \ln \frac{r}{r_0} - \frac{3}{2} \ln \left| 1 - \frac{r_0}{r} \right|\, . 
\end{align}
Then when $\frac{r_0}{r}<1$, 
\begin{align}
\label{eqf9}
\xi(r) = 2\xi_0 \left\{ \frac{1}{2} \ln \frac{1 + \sqrt{ 1 - \frac{r_0}{r}}}{1 - \sqrt{ 1 - \frac{r_0}{r}}} - \frac{1}{\sqrt{ 1 - \frac{r_0}{r}}} \right\} + \xi_1 \, ,
\end{align} 
Here $\xi_1$ is a constant of the integration. 
On the other hand, when $\frac{r_0}{r}>1$, one obtains
\begin{align}
\label{eqf9}
\xi(r) =2\xi_0 \left\{ \frac{i}{2} \ln \frac{1 + i \sqrt{ \frac{r_0}{r}-1}}{1 - i\sqrt{ \frac{r_0}{r}-1}} - \frac{1}{\sqrt{ \frac{r_0}{r} -1 }} \right\} + \xi_1 \, ,
\end{align} 
Here $\xi_1$ is also a constant of the integration. 
When $\xi_0=0$, the solution reduces to the standard Schwarzschild solution. 

\subsubsection{Non-trivial solution when $V\neq 0$}\label{SecIVA2}

We now consider a non-trivial solution when $V$ does not vanish. 

Let us now rewrite (\ref{eqf2}) as follows
\begin{align}
\label{eta}
0 =\e^{2\eta} \left\{ - \frac{1}{2} \left(\frac{1}{r} - \nu'\right) \left( \e^{-2\eta} \right)' + \left( \nu'' + {\nu'}^2 + \frac{\nu'}{r} - \frac{1}{r^2} \right) \e^{-2\eta}
+ \frac{1}{r^2} \right\} \, ,
\end{align}
which is an inhomogeneous linear equation for $\e^{-2\eta}$. 
By assuming $\e^{2\nu} = 1 - \frac{r_0}{r}$, we obtain, 
\begin{align}
\label{etaeq}
0 = - \frac{2 - \frac{3r_0}{r}}{4r \left(1 - \frac{r_0}{r} \right)} \left( \e^{-2\eta} \right)' 
+ \frac{- 4 + \frac{6r_0}{r} - \frac{3{r_0}^2}{r^2}}{4r^2\left(1 - \frac{r_0}{r}\right)^2} \e^{-2\eta}
+ \frac{1}{r^2} \, .
\end{align}
The Schwarzschild spacetime $\e^{-2\eta} = 1 - \frac{r_0}{r}$ must be solution of (\ref{etaeq}). 
The general solution of (\ref{etaeq}) is given by 
\begin{align}
\label{etasol}
\e^{-2\eta} = 1 - \frac{r_0}{r} + \frac{C_0 \left( r-r_0 \right)}{r \left( 2r - 3r_0 \right)^2}\, .
\end{align}
Here $C_0$ is a constant of the integration. 
Because $\e^{-2\eta}$ vanishes at the horizon $r=r_0$, there is no curvature singularity at the horizon but there could be a singularity at $r=\frac{3r_0}{2}$. 
The radius is larger than the horizon radius $r=r_0$, and therefore the singularity is naked. 
Even if we consider the limit of $r_0\to 0$, there remains non-trivial spacetime, 
\begin{align}
\label{etasollimit}
\e^{-2\eta} = 1 + \frac{C_0}{4r^2}\, ,
\end{align}
where the horizon vanishes and the singular surface is combined with the singularity 
at the origin, which is naked. 

By using (\ref{eqf1}), we obtain the solution of $\lambda$, 
\begin{align}
\label{eqf1eta1}
\lambda = \frac{2 \e^{-2\eta} \left( \nu' + \eta' \right) }{r} 
= \frac{4 C_0 \left( r-r_0 \right)r}{r^3 \left( 2r - 3r_0 \right)^3} \, .
\end{align}
By using (\ref{eq1}), we also find $V$ as a function of $r$
\begin{align}
\label{eq1etaV}
V= - 2 \e^{- 2 \eta} \left( \frac{2\eta'}{r} + \frac{\e^{2\eta} - 1}{r^2} \right) 
= \frac{2C_0 \left( -2r + r_0 \right)}{r^2 \left( 2r - 3r_0 \right)^3}\, .
\end{align}
One may choose $\omega(\phi)$ as in (\ref{ex4})
\begin{align}
\label{ex4eta }
\omega(\phi) = \e^{2\eta(r=\phi)} = \frac{1}{1 - \frac{r_0}{r} + \frac{C_0 \left( \phi -r_0 \right)}{\phi \left( 2\phi - 3r_0 \right)^2}} \, .
\end{align}
Then we may identify $\phi$ with the radial coordinate $r$. 
Then Eq.~(\ref{eq1etaV}) shows 
\begin{align}
\label{eq1etaV}
V(\phi) = \frac{2C_0 \left( -2\phi + r_0 \right)}{\phi^2 \left( 2\phi - 3r_0 \right)^3}\, .
\end{align}
Therefore we got a BH solution with non-trivial potential for the mimetic scalar $\phi$

\subsubsection{Black hole shadow in the model with $V\neq 0$}\label{SecIVA3}

The radius $r_\mathrm{sh}$ of the black hole shadow is given by the radius $r_\mathrm{ph}$ of the circular orbit of the photon which is called a photon sphere, as follows, 
\begin{align}
\label{shph}
r_\mathrm{sh}=\left. r\e^{-\nu(r)} \right|_{r=r_\mathrm{ph}}\, .
\end{align}
The following Lagrangian gives the motion of the photon, 
\begin{align}
\label{ph1g}
\mathcal{L}= \frac{1}{2} g_{\mu\nu} \dot q^\mu \dot q^\nu = \frac{1}{2} \left( - \e^{2\nu} {\dot t}^2 + \e^{2\eta} {\dot r}^2 + r^2 {\dot\theta}^2 + r^2 \sin^2 \theta {\dot\phi}^2 \right) \, .
\end{align}
Here the ``dot'' or ``$\dot\ $'' expresses the derivative with respect to the affine parameter. 
In the case of a photon, whose geodesic is null, we also require $\mathcal{L}=0$. 
Because the Lagrangian $\mathcal{L}$ does not depend on the $t$ and $\phi$, 
there are conserved quantities corresponding to energy $E$ and angular momentum $L$, 
\begin{align}
\label{phEg}
E \equiv&\, \frac{\partial \mathcal{L}}{\partial \dot t} = - \e^{2\nu} \dot t \, , \\
\label{phMg}
L \equiv&\, \frac{\partial V}{\partial\dot\phi}= r^2 \sin^2 \theta \dot\phi \, , 
\end{align}
We should also note that the total energy $\mathcal{E}$ of the system should be conserved, 
\begin{align}
\label{totalEg}
\mathcal{E} \equiv \mathcal{L} - \dot t \frac{\partial \mathcal{L}}{\partial \dot t} - \dot r \frac{\partial \mathcal{L}}{\partial \dot r} 
 - \dot\theta \frac{\partial \mathcal{L}}{\partial \dot\theta} - \dot\phi \frac{\partial \mathcal{L}}{\partial \dot\phi} = \mathcal{L} \, , 
\end{align}
Because we are considering the null geodesic, we require $\mathcal{E}=\mathcal{L}=0$. 
Without loss of generality, we consider the orbit on the equatorial plane with $\theta=\frac{\pi}{2}$. 
Then the condition $\mathcal{E}=\mathcal{L}=0$ gives, 
\begin{align}
\label{geo1g}
0= - \frac{E^2}{2} \e^{-2 \left( \nu + \eta\right)} + \frac{1}{2} {\dot r}^2 + \frac{L^2 \e^{- 2\eta}}{2r^2} \, ,
\end{align}
This system is analogous to the classical dynamical system with potential $U(r)$, 
\begin{align}
\label{geo2g}
0 =\frac{1}{2} {\dot r}^2 + U(r)\, , \quad U(r) \equiv \frac{L^2 \e^{- 2\eta}}{2r^2} - \frac{E^2}{2} \e^{-2 \left( \nu + \eta\right)}\, .
\end{align}
The radius of the circular orbit, where $\dot r=0$, is given by $U(r)= U'(r)=0$ by the analogy of classical mechanics. 
Then by using $\e^{2\nu} = 1 - \frac{r_0}{r}$ and (\ref{etasol}), we find
\begin{align}
\label{etaU1}
0 = U (r) = \frac{L^2}{2} \left( \frac{1}{r^2} - \frac{r_0}{r^3} + \frac{C_0 \left( r-r_0 \right)}{r^3 \left( 2r - 3r_0 \right)^2} \right)
 - \frac{E^2}{2} \left( 1 + \frac{C_0}{ \left( 2r - 3r_0 \right)^2} \right) \, .
\end{align}
and 
\begin{align}
\label{etaU2}
0 = U'(r) = \frac{L^2}{2} \left( - \frac{2}{r^3} + \frac{3r_0}{r^4} 
+ \frac{C_0\left( - 8 r^2 + 16 r_0 r - 9{r_0}^2 \right) }{r^4 \left( 2r - 3r_0 \right)^3} \right) + \frac{E^2 C_0}{ \left( 2r - 3r_0 \right)^3} \, .
\end{align}
The solution for $r$ of (\ref{etaU1}) and (\ref{etaU2}) is the radius of the photon sphere $r=r_\mathrm{ph}$. 
In the case of the standard Schwarzschild black hole, which corresponds to $C_0=0$, 
the solution is given by $E^2= - \frac{4L^2}{27{r_0}^2}$ and $r_\mathrm{ph}=\frac{3}{2}r_0$, 
which is 1.5 times larger than the horizon radius $r_\mathrm{horizon}=r_0$ as is well known. 

Eqs.~(\ref{etaU1}) and (\ref{etaU2}) give
\begin{align}
\label{etaV3}
0 = - \frac{1}{r^3} + \frac{3r_0}{2r^4} + \frac{C_0\left( - 10 r^2 + 26 r_0 r - 18 {r_0}^2 \right)}{2r^4 \left( 2r - 3r_0 \right)^3} 
+ \frac{{C_0}^2\left( - 6r^2 + 14 r_0 r - 9{r_0}^2 \right)}{2r^4 \left( 2r - 3r_0 \right)^5} \, , 
\end{align}
which can be rewritten as 
\begin{align}
\label{etaV4}
r= \frac{3r_0}{2} + \frac{C_0\left( - 10 r^2 + 26 r_0 r - 18 {r_0}^2 \right)}{2 \left( 2r - 3r_0 \right)^3} 
+ \frac{{C_0}^2\left( - 6r^2 + 14 r_0 r - 9{r_0}^2 \right)}{2 \left( 2r - 3r_0 \right)^5} \, .
\end{align}
This equation cannot be solved perturbatively with respect to $C_0$ because the solution when $C_0=0$ is the Schwarzschild solution $r=\frac{3r_0}{2}$ but 
the terms including $C_0$ has a pole at $r=\frac{3r_0}{2}$. 
We consider the possibility that a large photon sphere could appear. 
By assuming $r\gg r_0$, Eq.~(\ref{etaV4}) can be approximated as 
\begin{align}
\label{etaV4}
0 \sim r^4 + \frac{5C_0}{8} r^2 + \frac{3{C_0}^2}{32} = \left( r^2 + \frac{C_0}{4} \right) \left( r^2 + \frac{3C_0}{8} \right) \, ,
\end{align}
Therefore when $C_0<0$ and $-C_0\gg {r_0}^2$, there are solutions, 
\begin{align}
\label{etasol}
r = \frac{\sqrt{-C_0}}{2}\, , \ \frac{1}{2} \sqrt{ - \frac{3C_0}{2}}\, .
\end{align}
When $r\gg r_0$, Eq.~(\ref{etasol}) behaves as 
\begin{align}
\label{etasolas}
\e^{-2\eta} \sim 1 + \frac{C_0}{4r^2}\, .
\end{align}
When $r = \frac{\sqrt{-C_0}}{2}$, the r.h.s. of (\ref{etasolas}) vanishes and when $r = \frac{1}{2} \sqrt{ - \frac{3C_0}{2}}$, the r.h.s. is 
given by $1 - \frac{2}{3} = \frac{1}{3}>0$. 
Therefore the radius $r = \frac{1}{2} \sqrt{ - \frac{3C_0}{2}}$ could correspond to the radius of the photon sphere 
$r=r_\mathrm{ph} = \frac{1}{2} \sqrt{ - \frac{3C_0}{2}}$. 
As $\e^{-2\nu}\sim 1$ when $r\gg r_0$, the radius $r_\mathrm{sh}$ of the black hole shadow is almost identical with the radius $r_\mathrm{ph}$ 
of the photon sphere. 
Maybe we need more detailed check including the terms of $\mathcal{O}\left( \frac{r_0}{\sqrt{-C_0}} \right)$ in order to investigate if 
the radius $r = \frac{\sqrt{-C_0}}{2}$ corresponds to the radius of the photon sphere. 
Anyway, the singular surface $r=\frac{3r_0}{2}$ is hidden by the photon sphere and could not be observed by far observers.

The above result also shows that in the case $r_0=0$ in (\ref{etasollimit}), where the ADM mass $M\equiv \frac{r_0}{2}$ vanishes, there is a non-trivial object whose photon sphere 
has the radius of exactly $r_\mathrm{ph} = \frac{1}{2} \sqrt{ - \frac{3C_0}{2}}$, which is identical to the radius of the black hole 
shadow $r_\mathrm{sh}=r_\mathrm{ph}= \frac{1}{2} \sqrt{ - \frac{3C_0}{2}}$. 

In Ref.~\cite{Bambi:2019tjh}, it has been shown that for M87$^*$, the radius of the black hole shadow is limited to be $2r_\mathrm{sh}/M \sim 11.0\pm 1.5$ and in \cite{Vagnozzi:2022moj}, 
for Sgr A$^*$, $4.21\lesssim r_\mathrm{sh}/M \lesssim 5.56$. 
These bounds show that $r_\mathrm{sh}/r_0\sim 2.8$ for M87$^*$ and $2.1\lesssim r_\mathrm{sh}/r_0 \lesssim 2.8$ for Sgr A$^*$. 
Note in both cases, the singular surface at $r=1.5r_0$ is hidden by the black hole shadow. 

Eq.~(\ref{shph}) has the following form now, 
\begin{align}
\label{shph}
r_\mathrm{sh}= \frac{r_\mathrm{ph}}{\sqrt{ 1 - \frac{r_0}{r_\mathrm{ph}}}}\, .
\end{align}
By solving (\ref{shph}) with respect to $r_\mathrm{ph}$, we find 
\begin{align}
\label{rshph}
r_\mathrm{ph}= \left\{ \frac{ {r_\mathrm{sh}}^2 r_0 + \sqrt{ {r_\mathrm{sh}}^4 {r_0}^4 - \frac{4 {r_\mathrm{sh}}^6}{27}}}{2} \right\}^\frac{1}{3}
+ \left\{ \frac{ {r_\mathrm{sh}}^2 r_0 - \sqrt{ {r_\mathrm{sh}}^4 {r_0}^4 - \frac{4 {r_\mathrm{sh}}^6}{27}}}{2} \right\}^\frac{1}{3}
\end{align}
When $r_\mathrm{sh}/r_0\sim 2.8$, we find 
\begin{align}
\label{rshph2}
r_\mathrm{ph} \sim 3.26 r_0 \, ,
\end{align}
and when $r_\mathrm{sh}/r_0=2.1$ 
\begin{align}
\label{rshph2BB}
r_\mathrm{ph} \sim 2.49 r_0 \, ,
\end{align}
Both $r_\mathrm{ph}$ in (\ref{rshph2}) and (\ref{rshph2BB}) are larger than the radius $r=1.5r_0$ of the singular surface. 

By solving (\ref{etaV3}) with respect to $C_0$ after putting $r=r_\mathrm{ph}$, when $r_\mathrm{sh}/r_0\sim 2.8$, one obtains 
\begin{align}
\label{etaV3D}
C_0 \sim - 0.0973 {r_0}^2 \, , \quad - 0.00117 {r_0}^2 \, .
\end{align}
and when $r_\mathrm{sh}/r_0=2.1$, 
\begin{align}
\label{etaV3F}
C_0 \sim -3.93 {r_0}^2 \, , \quad -1.35 {r_0}^2 \, . 
\end{align}
Then in the case of M87$^*$, $C_0$ is given by (\ref{etaV3D}) and 
in the case of Sgr A$^*$, $-3.93 {r_0}^2 \lesssim C_0 \lesssim - 0.0973 {r_0}^2$ or $-1.35 {r_0}^2 \lesssim C_0 \lesssim - 0.00117 {r_0}^2$. 
In any case, there are solutions consistent with the observation. 

\subsection{Static and spherically symmetric spacetime in the scalar mimetic $F(R)$ gravity}\label{SecIVB}

In the scalar mimetic $F(R)$ gravity, 
the $(t,t)$, $(r,r)$, and $(i,j)$ components of Eq.~(\ref{aedenFR}) are given by 
\begin{align}
\label{eq1FR}
0=&\, \frac{1}{2}\e^{2\nu} F + \e^{2\left(\nu - \eta\right)} \left\{
\nu'' + \left(\nu' - \eta'\right)\nu' + \frac{2 \nu'}{r}\right\} F_R - \e^{2\left(\nu - \eta\right)} \left\{ F_R'' + \left( \frac{2}{r} - \eta' \right) F_R' \right\} 
\nonumber \\
&\, + \frac{1}{2}\e^{2\nu} V \, , 
\\
\label{eq2FR}
0=&\, - \frac{1}{2} \e^{2\eta} F + \left[ - \left\{ \nu'' + \left(\nu' - \eta'\right)\nu' \right\} + \frac{2 \eta'}{r} \right] F_R 
+ \left( \frac{2}{r} + \nu' \right) F_R' \nonumber \\
&\, - \frac{\e^{2\eta}}{2} V - \lambda \omega \left(\phi'\right)^2 \, , \\
\label{eq3FR}
0 =&\, - \frac{r^2}{2} F + \left[ 1 - \left\{1 + r \left(\nu' - \eta' \right)\right\}\e^{-2\eta}\right] F_R 
+ r^2 \e^{-2\eta} \left\{ F_R'' + \left( \frac{1}{r} + \nu' - \eta' \right) F_R' \right\} \nonumber \\
&\, - \frac{r^2}{2} V \, .
\end{align}
Other components vanish trivially. 

By combining (\ref{eq1FR}) and (\ref{eq2FR}), we obtain
\begin{align}
\label{eqFR1}
0 \frac{2\nu' + 2 \eta'}{r} F_R - F_R'' + \left( \nu' + \eta' \right) F_R' - \lambda \e^{2\eta} \, .
\end{align}
Here we have used (\ref{eq4}). 
On the other hand, by combining (\ref{eq1FR}) and (\ref{eq3FR}), one finds
\begin{align}
\label{eqFR2}
0= \left[ \e^{- 2 \eta} \left\{
\nu'' + \left(\nu' - \eta'\right)\nu' + \frac{\nu'+ \eta'}{r} - \frac{1}{r^2} \right\} 
+ \frac{1}{r^2} \right] F_R + \e^{- 2 \eta} \nu' F_R' \, ,
\end{align}
If $\nu=\nu(r)$ and $\eta=\eta(r)$ are given, one obtains the $r$ dependence of the $F_R$ is found by 
\begin{align}
\label{FRsol1}
F_R=F_R(r) = F_0 \exp \left( - \int \frac{dr}{\nu'} \left\{ \nu'' + \left(\nu' - \eta'\right)\nu' + \frac{\nu'+ \eta'}{r} - \frac{1}{r^2} + \frac{\e^{2\eta}}{r^2} \right\} \right)\, .
\end{align}
Then we can solve (\ref{eqFR1}) with respect to $\lambda$, $\lambda=\lambda(r)$. 

By using $\nu=\nu(r)$ and $\eta=\eta(r)$, we also find the $r$-dependence of the scalar curvature $R$ by using the last expression in (\ref{GBv}), $R=R(r)$, which could be 
solved with respect to $r$, $r=r(R)$. 
By substituting the expression $r=r(R)$, we find the $R$-dependence of $F_R$, $F_R(R)=F_R \left(r=r \left(R\right) \right)$. 
By integrating $F_R(R)$ with respect to $R$, we obtain the form of $F(R)$, $F(R)=\int dR F_R(R)$. 
Then by using (\ref{eq1FR}) or (\ref{eq3FR}), it is found the $r$-dependence of $V$, $V=V(r)$. 
We may solve Eq.~(\ref{eq4}) with respect the mimetic scalar field $\phi$, $\phi=\phi(r)$. 
We may algebraically solve the obtained expression $\phi=\phi(r)$ with respect to $r$, $r=r(\phi)$. 
By substituting the expression of $V=V(r)$ obtained from (\ref{eq4}), we find $V$ as a function of $\phi$, $V=V(\phi)=V\left( r=r\left(\phi\right) \right)$. 
Therefore for a given geometry expressed by $\nu=\nu(r)$ and $\eta=\eta(r)$, 
we can obtain the model realising the BH geometry by adjusting the functional forms of $F(R)$ and $V(\phi)$. 

\subsection{Hayward black hole in scalar mimetic $F(R)$ gravity}\label{SecIVC}

As an example, we may consider the Hayward black hole~\cite{Hayward:2005gi}, 
\begin{align}
\label{Hay1}
\e^{2\nu}=\e^{-2\eta}= 1 - \frac{r_0 r^2}{r^3 + r_0 \lambda^2}\, .
\end{align}
Here $\lambda$ is a parameter with the dimension of the length and $M=\frac{r_0}{2}$ corresponds to the ADM mass. 

\subsubsection{Properties of Hayward black hole}\label{SecIVC1}

We now briefly review the properties of the Hayward black hole in (\ref{Hay1}). 

First, we should note that $\left(\e^{2\nu} \right)'$ vanishes at the centre and therefore, there is no conical singularity or any other kind of singularity. 
When $r$ is small, $\e^{2\nu}$ behaves as $\e^{2\nu} \sim 1 - \frac{r^2}{\lambda^2}$, that is, the spacetime becomes the asymptotically de Sitter spacetime.

When we rewrite $\e^{2\nu}$ in (\ref{Hay1}) as follows,
\begin{align}
\label{Hayward2}
\e^{2\nu} = \frac{b(r)}{r^3 + r_0\lambda^2} \, , \qquad \qquad b(r)\equiv r^3 - r_0 r^2 + r_0 \lambda^2 \, ,
\end{align}
the solutions of the equation $b'(r) = 3r^2 - 2r_0 r=0$ are given by $r=0$ and $r=\frac{2}{3}r_0$. 
The equation 
\begin{align}
\label{Hayward3}
b\left( r =\frac{2}{3}r_0 \right) = - \frac{2^2}{3^3} {r_0}^3 + r_0 \lambda^2 \, ,
\end{align}
gives the following, 
\begin{enumerate}
\item If $\frac{2^\frac{2}{3}r_0}{3\left( r_0 \lambda^2 \right)^\frac{1}{3}}<1$, 
$\e^{2\nu}$ does not vanish and positive. 
Therefore the spacetime given by (\ref{Hay1}) is a kind of the gravasar~\cite{Mazur:2001fv}. 
\item When $\frac{2^\frac{2}{3}r_0}{3\left( r_0 \lambda^2 \right)^\frac{1}{3}}>1$, 
$\e^{2\nu}$ vanishes twice corresponding to the outer and inner horizons. 
\item In the case $\frac{2^\frac{5}{3}M}{3\left( 2M\lambda^2 \right)^\frac{1}{3}}=1$, 
the radii of the two horizons coincide with each other corresponding to the extremal black hole. 
\end{enumerate}

\subsubsection{Construction of a model realizing Hayward black hole}\label{SecIVC2}

We now construct a model realizing the Hayward black hole in (\ref{Hay1}). 

By using (\ref{FRsol1}), we find 
\begin{align}
\label{FRsol1Hay}
F_R(r) =&\, F_0 \exp \left( - \int \frac{dr}{\left(\e^{2\nu} \right)'} \left\{ \left( \e^{2\nu} \right)'' - \frac{\e^{2\nu}}{r^2} + \frac{1}{r^2} \right\} \right) \nonumber \\
=&\, F_0 \left( r^3 - 2r_0 \lambda^2 \right)^{-\frac{3}{2}} \left( r^3 + r_0 \lambda^2 \right)^2 \, .
\end{align}
Here $F_0$ is a constant of the integration. 
On the other hand, the scalar curvature $R$ is given by 
\begin{align}
\label{HayR}
R= \frac{3r_0 r^6 - 12 {r_0}^2 \lambda^2 r^3 + 3 {r_0}^3 \lambda^4 }{\left(r^3 + r_0 \lambda^2\right)^3} \, , 
\end{align}
which can be solved with respect to $r$, as follows, 
\begin{align}
\label{r1}
r^3 + r_0 \left( \lambda^2 - \frac{1}{R} \right) = \alpha_+ + \alpha_- \, , \ \zeta \alpha_+ + \zeta^2 \alpha_- \, , \ \zeta^2 \alpha_+ + \zeta \alpha_- \, .
\end{align}
Here $\zeta\equiv \e^{\frac{2\pi}{3}i} = - \frac{1}{2} + \frac{\sqrt{3}}{2} i$ and 
\begin{align}
\label{alphapm}
\alpha_\pm = \left\{ \frac{9 {r_0}^3 \lambda^4}{R} - \frac{9 {r_0}^3 \lambda^2}{R^2} + \frac{{r_0}^3}{R^3} \pm \sqrt{ 
\left( \frac{9 {r_0}^3 \lambda^4}{R} - \frac{9 {r_0}^3 \lambda^2}{R^2} + \frac{{r_0}^3}{R^3} \right)^2 + \frac{125 {r_0}^6 \lambda^6}{R^3}
} \right\}^\frac{1}{3} \, .
\end{align}
Because $\alpha_\pm$ are real numbers, in order that $r$ is a real number, the solution is given by 
\begin{align}
\label{r2}
r^3 + r_0 \left( \lambda^2 - \frac{1}{R} \right) = \alpha_+ + \alpha_- \, .
\end{align}
Then by combining the expressions of (\ref{FRsol1Hay}) and (\ref{r2}), we find the explicit form of $F(R)$. 
also 
By using (\ref{eqFR1}), we find the solution of the Lagrange multiplier field $\lambda$ as a function of $r$, 
\begin{align}
\label{lambda}
\lambda =&\, - \e^{2\nu} F_R'' = - F_0 \left( 1 - \frac{r_0 r^2}{r^3 + r_0 \lambda^2} \right) \frac{59 r_0 \lambda^2 r^9 + 466 {r_0}^2 \lambda^4 r^6 
+ 71 {r_0}^3 \lambda^6 r^3 + 6 {r_0}^4 \lambda^8 }{2 r^2 \left( r^3 - 2r_0 \lambda^2 \right)^\frac{7}{4}} \, .
\end{align}
We also find $F(R)$ by integrating $F_R$ with respect to $R$, 
\begin{align}
\label{F}
F =&\, \int dR F_R = \int dr \frac{dR}{dr} F_R \nonumber \\
=&\, \frac{2}{3} r_0 F_0 \left[ - \frac{1}{\sqrt{r^3 - 4M\lambda^2}} + \frac{1}{\sqrt{3r_0\lambda^2}} \left\{ 
 - \frac{35}{2} \mathrm{Arctan} \left( \sqrt{\frac{3r_0 \lambda^2}{r^3 - 2r_0 \lambda^2}} \right) \right. \right. \nonumber \\
&\, \left. \left. \qquad \qquad + \frac{19}{6} \frac{\sqrt{\frac{3r_0 \lambda^2}{r^3 - 2r_0 \lambda^2}}}
{\frac{3r_0 \lambda^2}{r^3 - 2r_0 \lambda^2} + 1}\right\} \right] + F_1 \, .
\end{align}
Here $F_1$ is a constant of the integration. 
As we will see soon, $F_1$ is an irrelevant parameter. 
By substituting (\ref{r2}) into (\ref{F}), we find the explicit functional form of $F(R)$. 

The potential $V$ for the mimetic scalar $\phi$ can be found by using (\ref{eq3FR}), as follows, 
\begin{align}
\label{V}
V(r) =&\, \, - F + \frac{2}{r^2} \left[ 1 - \left\{1 + r \left(\nu' - \eta' \right)\right\}\e^{-2\eta}\right] F_R 
+ 2 \e^{-2\eta} \left\{ F_R'' + \left( \frac{1}{r} + \nu' - \eta' \right) F_R' \right\} \nonumber \\
=&\, \, - F + \frac{F_R} {r^2\left(r^3 + r_0 \lambda^2\right)^3 \left( r^3 - 2r_0 \lambda^2 \right)^2}
\left[ {r_0}^2 \lambda^2 r^2 \left( - 59 r^9 - 376 r_0 \lambda^2 r^6 - 219 {r_0}^2 \lambda^4 r^3 \right. \right. \nonumber \\
&\, \left. + 30 {r_0}^3 \lambda^6 \right) + 2 r^{15} + 27 r_0 \lambda^2 r^{12} + 521 {r_0}^2 \lambda^4 r^9 
+ 629 {r_0}^3 \lambda^6 r^6 + 135 {r_0}^4 \lambda^8 r^3 \nonumber \\
&\, \left. + 14 {r_0}^5 \lambda^{10} \right] \, .
\end{align}
As clear from the expression in (\ref{V}), $V$ includes the parameter $F_1$ in (\ref{F}) via the term $-F$ as $-F_1$. 
As clear from Eqs.~(\ref{eq1FR}), (\ref{eq2FR}), and (\ref{eq3FR}), $V$ and $F$ appear only in the combination of of the sum $V+F$. 
Therefore $- F_1$ in $V$ is cancelled by $F_1$ in $F$ and therefore $F_1$ does not appear in the equations. 

We now choose $\omega(\phi)$ as in (\ref{ex4})
\begin{align}
\label{ex4}
\omega(\phi) = \e^{2\eta(r=\phi)} = \frac{1}{1 - \frac{r_0\phi^2}{\phi^3 + r_0 \lambda^2}} \, .
\end{align}
Then we may identify $\phi$ with $r$, $\phi=r$ and $V(\phi)$ is obtained from (\ref{V}) by $V(\phi)=V(r=\phi)$. 

In summary, we constructed an explicit model of the scalar mimetic $F(R)$ gravity realising the Hayward black hole. 

\subsubsection{Photon orbit and the radius of the black hole shadow}\label{SecIVC3}

Let us now consider the black hole shadow in the Hayward black hole (\ref{Hay1}). 
The equations corresponding to 
Eqs.~(\ref{ph1g}), (\ref{phEg}), and (\ref{phMg}) have the following forms
\begin{align}
\label{ph1}
\mathcal{L}=&\, \frac{1}{2} g_{\mu\nu} \dot q^\mu \dot q^\nu = \frac{1}{2} \left( - \e^{2\nu} {\dot t}^2 + \e^{-2\nu} {\dot r}^2 
+ r^2 {\dot\theta}^2 + r^2 \sin^2 \theta {\dot\phi}^2 \right) \, , \\
\label{phE}
E \equiv&\, \frac{\partial \mathcal{L}}{\partial \dot t} = - \e^{2\nu} \dot t \, , \\
\label{phM}
L \equiv&\, \frac{\partial \mathcal{L}}{\partial\dot\phi}= r^2 \sin^2 \theta \dot\phi \, , 
\end{align}
The condition $\mathcal{L}=0$ gives an analogue of the classical dynamical system with potential $U(r)$, 
\begin{align}
\label{geo2}
0=\frac{1}{2} {\dot r}^2 + U(r)\, , \quad U(r) \equiv \frac{L^2 \e^{2\nu(r)}}{2r^2} - \frac{E^2}{2}\, .
\end{align}
The radius of the circular orbit is again given by $U(r)=U'(r)=0$ by the analogy with classical mechanics. 
By using (\ref{Hay1}), we find
\begin{align}
\label{HayU}
0 = U (r) = \frac{L^2}{2} \left( \frac{1}{r^2} - \frac{r_0}{r^3 + r_0\lambda^2} \right) - \frac{E^2}{2} \, ,
\end{align}
and 
\begin{align}
\label{HayU2}
0 = U'(r) = L^2 \left( - \frac{1}{r^3} + \frac{3r_0r^2}{2\left(r^3 + r_0 \lambda^2\right)^2} \right)\, ,
\end{align}
Let the radius of the photon sphere be $r=r_\mathrm{ph}$. 
Then Eq.~(\ref{HayU2}) can be rewritten as 
\begin{align}
\label{HayU3}
r_\mathrm{ph} = \frac{3r_0}{2} \frac{r_\mathrm{ph}^6}{\left({r_\mathrm{ph}}^3 + r_0 \lambda^2\right)^2} < \frac{3}{2}r_0 \, .
\end{align}
We should note $\frac{r_\mathrm{ph}^6}{\left({r_\mathrm{ph}}^3 + r_0 \lambda^2\right)^2}<1$. 
Therefore although it is difficult to solve (\ref{HayU3}) with respect to $r_\mathrm{ph}$, 
if the ADM mass $M\equiv \frac{r_0}{2}$ is fixed, 
the radius of the photon sphere becomes smaller compared with the case of the Schwarschild black hole, 
where the radius of the photon sphere is given by $r=\frac{3}{2}r_0$. 
We should note that Eq.~(\ref{geo2}) determine $E$ as a function of $r_0$ and $L$. 

The radius $r_\mathrm{sh}$ of the black hole shadow is given by using (\ref{shph}) as follows, 
\begin{align}
\label{Haysh}
r_\mathrm{sh} = \frac{r_\mathrm{ph}}{\sqrt{1 - \frac{r_0 {r_\mathrm{ph}}^2}{{r_\mathrm{ph}}^3 + r_0 \lambda^2}}} 
= \frac{3\sqrt{3} r_0}{2} \frac{r_\mathrm{ph}^6}{\left({r_\mathrm{ph}}^3 + r_0 \lambda^2\right)^2} 
\frac{\frac{1}{\sqrt{3}}}{\sqrt{1 - \frac{r_0 {r_\mathrm{ph}}^2}{{r_\mathrm{ph}}^3 + r_0 \lambda^2}}} \, .
\end{align}
Because 
\begin{align}
\label{Haysh2}
\frac{\frac{1}{\sqrt{3}}}{\sqrt{1 - \frac{r_0 {r_\mathrm{ph}}^2}{{r_\mathrm{ph}}^3 + r_0 \lambda^2}}} 
= \frac{ \sqrt{ 1 - \frac{r_0}{\frac{3}{2}r_0}}}{\sqrt{1 - \frac{r_0 {r_\mathrm{ph}}^2}{{r_\mathrm{ph}}^3 + r_0 \lambda^2}}} 
< \frac{ \sqrt{ 1 - \frac{r_0}{r_\mathrm{ph}}}}{\sqrt{1 - \frac{r_0 {r_\mathrm{ph}}^2}{{r_\mathrm{ph}}^3 + r_0 \lambda^2}}} <1 \, ,
\end{align}
and $\frac{r_\mathrm{ph}^6}{\left({r_\mathrm{ph}}^3 + r_0 \lambda^2\right)^2}<1$, we find 
\begin{align}
\label{Haysh}
r_\mathrm{sh} = \frac{r_\mathrm{ph}}{\sqrt{1 - \frac{r_0 {r_\mathrm{ph}}^2}{{r_\mathrm{ph}}^3 + r_0 \lambda^2}}} 
< r_\mathrm{sh} = \frac{3\sqrt{3} r_0}{2} \, .
\end{align}
Therefore the radius of the black hole shadow is also smaller than that of the standard Schwarzschild black hole, 
which is $\frac{3\sqrt{3} r_0}{2}$. 

For M87$^*$, we find $r_\mathrm{sh} r_0\sim 2.8 r_0$~\cite{Bambi:2019tjh} and 
for Sgr A$^*$, $2.1 r_0 \lesssim r_\mathrm{sh} \lesssim 2.8 r_0$ for Sgr A$^*$. 
The radius $r_\mathrm{sh} = \frac{3\sqrt{3} r_0}{2}\sim 2.598 r_0$ for the standard Schwarzschild black hole is consistent with the observations. 
We should note that there is also a possibility that Sgr A$^*$ might be the Hayward black hole. 

\section{Conclusion}\label{SecV}

In the case of the standard mimetic theory~\cite{Chamseddine:2013kea}, there exists the Schwarzschild black hole solution although the mimetic scalar field 
is time-dependent~\cite{Gorji:2020ten}. 
In our model, by the prescription in (\ref{cnstrnt2}) proposed in \cite{Nojiri:2022cah}, even for the Schwarzschild solution, the mimetic scalar field can be static. 
It has been shown in Ref.~\cite{Gorji:2020ten} that there are several kinds of spherically symmetric solutions, including the spacetime with a naked singularity 
and the black hole spacetime. 
Recently in \cite{Khodadi:2024ubi}, the radius of the photon sphere in these models has been investigated in order to compare these solutions with 
the observations by the Event Horizon Telescope~\cite{EventHorizonTelescope:2019dse}. It has been found that these solutions are excluded by the observations.
In other words, due to the lack of compact objects, the simplest mimetic gravity does not seem to be a realistic theory. 

In this paper, we have constructed solutions of the spherically symmetric and static spacetime by introducing the potential for the mimetic scalar field as well as modifying 
the gravitational sector for $F(R)$ theory. 
For the solution (\ref{etasol}), there appears the singular surface but this solution can have a large radius of the black hole shadow and therefore the singular surface should not be 
seen by far observers. 
Furthermore, even if the ADM mass vanishes in this model, there appears a photon sphere, which might be found by the observations as in the Event Horizon Telescope. 
We also proposed an $F(R)$ gravity extension of the scalar mimetic gravity (\ref{actionmimeticfractionFR}) as in \cite{Nojiri:2014zqa}. 
In the framework of the scalar mimetic $F(R)$ gravity, we have constructed a model, where the Hayward black hole~\cite{Hayward:2005gi} is a solution. 
The Hayward black hole is a regular black hole with two horizons. 
The radius of the black hole shadow becomes smaller compared with that in the Schwarzschild black hole with the same ADM mass. 
Therefore, the radius might be observed in future observations. Then these versions of scalar mimetic gravity seem to be consistent with 
the Event Horizon Telescope observations.

The inflation and dark energy cosmologies for scalar mimetic $F(R)$ gravity are also constructed. 
The confrontation of the theory with observational bounds 
may reduce the freedom in the choice of functional forms of scalar potential and $F(R)$ as well as give precise constraints to theory parameters. 
This will be investigated elsewhere. 

\section*{ACKNOWLEDGEMENTS}

This work was partially supported by the program Unidad de Excelencia Maria de Maeztu CEX2020-001058-M, Spain (S.D.O).

\end{document}